\input harvmac
\input epsf

\Title{\vbox{\baselineskip12pt\hbox{KUCP-0163}\hbox{{\tt hep-ph/0007290}}}}
{\centerline{Gravitational Stability and Extra Timelike Dimensions}}

\centerline{Satoshi Matsuda\footnote{$^\dagger$}{matsuda@phys.h.kyoto-u.ac.jp}}
{\it \centerline{Department of Fundamental Sciences, FIHS}
\centerline{Kyoto University, Kyoto 606-8501, Japan}}
\medskip
\centerline{Shigenori Seki\footnote{$^\ddagger$}{seki@phys.h.kyoto-u.ac.jp} }
{\it \centerline{Graduate School of Human and Environmental Studies}
\centerline{Kyoto University, Kyoto 606-8501, Japan}}

\vskip .3in

\centerline{{\bf abstract}}

Tachyonic Kaluza-Klein modes normally appear from the compactification of extra 
timelike dimensions with scale $L$ and lead to some instabilities of physical systems. 
We calculate the contribution of tachyonic gravitons to the self energies of 
spherical massive bodies of radius $R$ and discuss their possible gravitational stability.
We find that some spherical bodies can be stable at critical radii $R=2\pi L p$ 
for some positive integers $p$.
We also prove the generic property of massive bodies that 
for the range $0<R\leq\pi L$ the gravitational force due to the ordinary massless graviton exchange 
is screened by the Kaluza-Klein mode exchange of tachyonic gravitons.

\Date{July 2000}

\newsec{Introduction}
Recently large extra dimensions, which are normally spacelike ones,
are discussed. In 
\ref\ADD{N. Arkani-Hamed, S. Dimopoulos and G. Dvali, ``Phenomenology, astrophysics and cosmology 
of theories with sub-millimeter dimensions and TeV scale quantum gravity", 
Phys. Rev. D {\bf 59} (1999) 086004-1, {\tt hep-ph/9807344}; 
I. Antoniadis, S. Dimopoulos and G. Dvali, 
``Millimeter-range forces in superstring theories with weak-scale compactification", 
Nucl. Phys. {\bf B516} (1998) 70, {\tt hep-ph/9710204}.}
it is argued that 
the large scale of extra spacelike dimensions has a size of, or smaller than, a millimeter range  
in order not to be in conflict with any experiments and observations. 
On the other hand, extra timelike dimensions can also be considered
\ref\DGS{G. Dvali, G. Gabadadze and G. Senjanovi{\' c}, ``Constraints on extra time dimensions'', 
{\tt hep-ph/9910207}.}.

By compactifying extra dimensions, we find that Kaluza-Klein
modes are produced. They are non-tachyonic when the extra dimensions are 
spacelike, but tachyonic Kaluza-Klein modes are obtained when the extra 
dimensions are timelike. Now for simplicity we consider one extra timelike
dimension and compactify it on a circle of radius $L$ 
\ref\Y{F. J. Yndur{\' a}in, ``Disappearance of matter due to causality and 
probability violations in theories with extra timelike dimensions", Phys. Lett. {\bf B256} (1991) 15.}. 
Let gravitons propagate in the extra dimension, then  
we obtain tachyonic gravitons of the Kaluza-Klein modes. 
Since their propagators are proportional to 
$$
-i{1 \over k_0^2-{\bf k}^2+ {n^2 \over L^2} + i \epsilon},\quad n \in {\bf Z},
$$
up to a spin tensor factor, we can calculate the gravitational potential between two unit mass points 
at distance $d$ as  
\eqn\eqgp{
V(d) = -G_N{1 \over d} - \sum_{n = -\infty, n \neq 0}^{\infty} G_N {1 \over d}
\exp{\left( i{|n|\over L}d \right)},
}
in the nonrelativistic tree-level approximation, where $G_N$ is the Newton constant.
The first term of the potential \eqgp\ is the contribution of ordinary massless graviton
and the second is the one of tachyonic gravitons. 

Since the gravitational potential \eqgp\  has an imaginary part, 
which violates causality and conserved probability, it makes massive bodies unstable.
But the effect is not unacceptable  experimentally if the scale $L$ of 
the extra dimension is bounded below a sufficiently small size \DGS\Y.

In this paper we calculate the gravitational self energies of spherical massive bodies 
exactly and 
discuss their stability on an analytic ground. 
In section 2 we assume their mass densities as some typical ones and calculate 
the corresponding self energies.  
In section 3 we devote ourselves to discussions on the gravitational screening effect of 
tachyonic gravitons \DGS\ and  
about the gravitational stability of spherical bodies.
In the final section we present conclusions with some comments. 
We shall give some useful formulas in the appendix.

\newsec{The self energy of spherical body}
We consider the self energy of a spherical body of radius $R$ with mass 
density $\rho(r)$, which depends on the radial coordinate $r$.
Using the form of the potential given in the second term of  Eq.\eqgp, and 
integrating out the relative polar angles between two mass points of densities $\rho(r)$ 
and $\rho(l)$  
at distance $d=\sqrt{r^2+l^2-2rl\cos\beta}$ with their opening angle $\beta$, we obtain 
the contribution of the $n$-th Kaluza-Klein mode, as a tachyonic graviton, 
to the self energy as follows:
\eqn\eqt{
E_n = 8 i \pi^2 G_N L \int_0^R dr \int_0^r dl\ \rho(r) \rho(l){ r l \over |n|}  
\left[ \exp\left({i |n| \over L}(r + l)\right) 
- \exp\left({i |n| \over L}(r - l)\right)\right],
}
whereas 
from the first term of Eq.\eqgp\ 
the contribution of an ordinary massless graviton to that 
is given by 
\eqn\eqo{
E_0 = - 16 \pi^2 G_N \int_0^R dr \int_0^r dl\ \rho(r) \rho(l) r l^2 .
}
Note that this can also be obtained simply as the limit of $|n|\to 0$ from Eq.\eqt.
So we obtain the total gravitational self energy as
\eqn\eqene{
E(R) = E_0+\sum_{n = -\infty,n\neq 0}^{\infty}E_n = E_0 + 2 \sum_{n = 1}^{\infty}E_n .
}

In order to make definite and precise arguments on the implications of the tachyonic 
Kaluza-Klein modes to 
gravitational stability, we  actually set $\rho(r)$ to a few typical forms of mass density and 
calculate the corresponding gravitational self energies as follows.

\subsec{The $\rho(r) = {D \over r}$ case}
At first we set the density $\rho(r) = {D \over r}$, where $D$ is constant. 
This provides a simple and interesting result. From Eq.\eqo\ the contribution
of massless graviton to the self energy becomes
\eqn\eqoi{E_0 = - 16 \pi^2 G_n D^2 \int_0^R dr \int_0^r dl\ l  
      = - {8\pi^2 \over 3}G_N D^2 R^3 ,
}
and from Eq.\eqt\ the contribution of tachyonic gravitons becomes
\eqn\eqti{
2 \sum_{n = 1}^{\infty} E_n = 16 i \pi^2 G_N L D^2 
\sum_{n = 1}^{\infty}{1 \over n} \int_0^R dr \int_0^r dl\  f_n(r,l) , }
where
\eqn\fn{
 f_n(r,l) \equiv \exp\left({i n \over L}(r + l)\right)
- \exp\left({i n \over L}(r - l)\right) .}
We set $R = 2\pi L k + c$, where $k \in \{{\bf N},0\}$ and $0\leq c < 2 \pi L$. 
From Eqs.\eqene, \eqoi\ and \eqti\ the total self energy is given by
\eqn\eqenei{
E(R) = - {8\pi^2 \over 3}G_N D^2 (R^3 - 2\pi^3L^3k) +
16i\pi^2G_NLD^2\sum_{n = 1}^{\infty}{1 \over n}\int_0^c dr \int_0^r dl \ f_n(r,l) .
}
At $R = 2\pi L k$\ $(c = 0)$, we obtain
\eqn\enecr{
E(2\pi L k) = - {8\pi^2 \over 3}G_N D^2 2(\pi L)^3 k (4k^2 - 1) ,
}
and this is pure real. So we find that the spherical body with a critical radius $R=2\pi L k$ is 
gravitationally stable.

From Eqs.\eqenei\ and (A.2) the imaginary part of the self energy can be expressed as 
\eqnn\eqeneii
$$\eqalignno{
\Im E(2 \pi L k + c) &=  16 \pi^2 G_N D^2 L \int_0^c dr \int_0^r dl\ 
\sum_{n=1}^\infty {1 \over n}
\left(\cos {n(r + l) \over L} - \cos {n(r - l) \over L}\right) \cr
&= - 16 \pi^2 G_N D^2 L \int_0^c dr \int_0^r dl\  
\log {\left| \sin{r + l \over 2L}\right| \over \sin{r - l \over 2L}}, &\eqeneii 
}$$
where $0 \leq c < 2 \pi L$. Eq.\eqeneii\ exhibits a periodicity of the imaginary part 
with respect to the size $R$ of spherical body 
and the period $2\pi L$ is determined 
by the scale of the extra timelike dimension. 
We can calculate Eq.\eqeneii\ numerically and the result is presented in fig.1 .
\bigskip
\vbox{
\centerline{\epsfbox{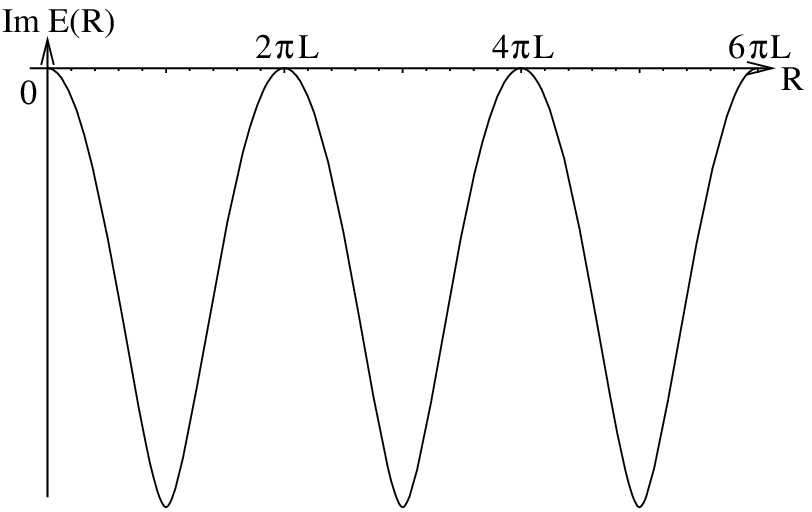}}
\vskip .5cm
\centerline{\fig\figi{The imaginary part of the self energy.}\ 
The imaginary part of the self energy $(\rho = {D \over r})$ .}}
\bigskip

From Eqs.\eqenei\ and (A.1) the real part of the self energy can be calculated as 
\eqnn\eqeneiii
$$\eqalignno{
  &\Re E(2 \pi L k + c) = - {8\pi^2 \over 3}G_N D^2 (R^3 - 2\pi^3L^3k)  \cr
  &  \qquad\qquad\qquad\qquad\qquad    - 16\pi^2G_N D^2 L\int_0^c dr \int_0^r dl\  
\sum_{n=1}^\infty {1 \over n}
\left(\sin {n(r + l) \over L} - \sin {n(r - l) \over L}\right) \cr
  &= - 32 \pi^3G_N D^2 L \cases{
k\left[3c^2 + 6 \pi L k c + (4k^2 - 1)\pi^2 L^2 \right] ,
&if $0 \leq c < \pi L, $  \cr
(k + 1)\left[3 c^2 + 6(k - 1)\pi L c + (4k^2 -4k + 3)\pi^2 L^2\right], 
&if $\pi L \leq c < 2 \pi L. $    \cr} \qquad \qquad &\eqeneiii}
$$
The behavior of Eq.\eqeneiii\ is shown in fig.2 .
\bigskip
\vbox{
\centerline{\epsfbox{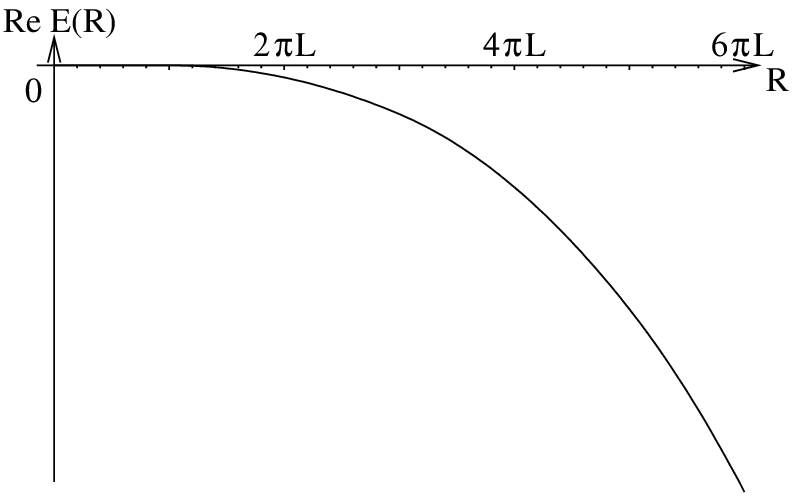}}
\vskip .5cm
\centerline{\fig\figii{The real part of the self energy.}\ 
The real part of the self energy $(\rho = {D \over r})$ .}}
Since Eq.\eqeneiii\ indicates that the real part of the self energy vanishes identically 
for  the range $0 < R \leq \pi L$ with $k=0$, 
we find that the gravitational force 
due to the ordinary massless graviton exchange 
is ``screened" in this region by the exchange of the Kaluza-Klein mode tower of 
tachyonic gravitons.

\subsec{The $\rho(r) = D_0$(constant) case}
We set $\rho(r)$ to a constant value $D_0$. This is also a physically normal 
and simple situation. Using Eq.\eqo\ the contribution of an ordinary graviton 
to the self energy of a spherical body of radius $R$ becomes
\eqn\eqconso
{
E_0 = - 16 \pi^2 G_N D_0^2 \int_O^R dr \int_0^r dl\  r l^2 
    = - {16 \over 15} \pi^2 G_N D_0^2 R^5 , 
}
and the one of tachyonic gravitons to that becomes
\eqn\eqconst{
2\sum_{n = 1}^\infty E_n =  16 i \pi^2 G_N L D_0^2 \sum_{n = 1}^\infty {1 \over n}
\int_0^R dr \int_0^r dl\  r l f_n(r,l) .}
At $R = 2 \pi L k $ with  $k \in \{{\bf N},0\}$, we obtain
\eqn\eqconsk{
E(2\pi L k) = - {16 \over 45}\pi^7 G_N D_0^2 L^5 k(4k^2-1)(24k^2+1) 
- 64 i G_N D_0^2 \pi^4 L^5 k^2 \zeta(3), 
}
where $\zeta$ is a zeta-function 
and $\zeta(3) = \sum_1^\infty {1 \over n^3} \sim {\pi^2 \over 25.79436}$. 
Since the imaginary part of Eq.\eqconsk\ does not vanish for any $k > 0$,
at the points $R=2\pi L k$ the spherical bodies are unstable and this result 
is quite different from the one in the case of $\rho(r) = {D \over r}$.

From Eqs.\eqene , \eqconso\ and \eqconst\ we calculate the imaginary part of 
the self energy as
$$
\Im E(R) = - 16 \pi^2 G_N D_0^2 L \int_0^R dr \int_0^r dl\  rl 
\log \left| {\sin{r + l \over 2L} \over \sin{r - l \over 2L}}\right|,
$$
and its numerical result is as shown in fig.3 .
\bigskip
\vbox{
\centerline{\epsfbox{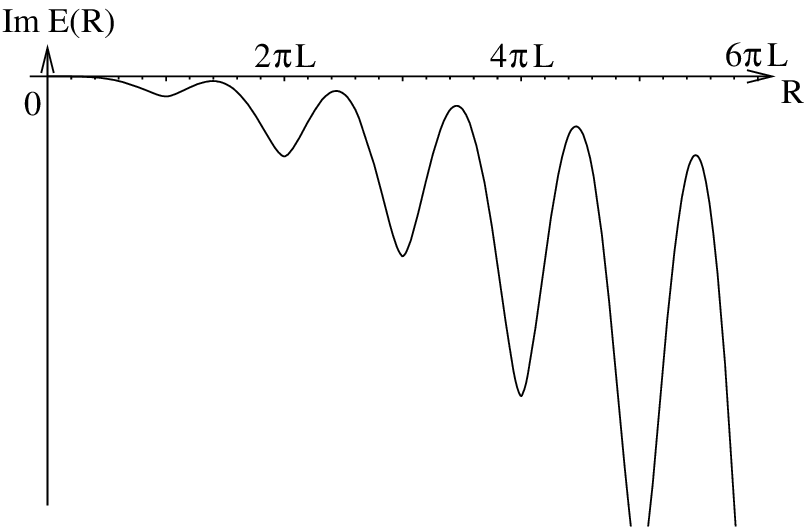}}
\vskip .5cm
\centerline{\fig\figiii{The imaginary part of the self energy.}\ 
The imaginary part of the self energy $(\rho = D_0)$ .}}
\bigskip\noindent
From fig.3 we note that the imaginary part of self energy never vanishes for any $R > 0$.
This implies that in the present case of mass density there is no critical value of radius $R$ 
which stabilizes the corresponding spherical body.

Now we set $R = 2\pi Lk + c$, where $k \in \{{\bf N}, 0\}$ and $0 \leq c < 2\pi L$. 
From Eqs.\eqene , \eqconso\ and \eqconst\ the real part of the self energy becomes
at $0 \leq c < \pi L$
\eqn\eqconsi{
\eqalign{
\Re E(2 \pi L k + c) = &- {16 \over 45} G_N D_0^2 L^2 \pi^4 (2k + 1)k
\left[30 c^3 + 15 (8k -1)\pi L c^2\right. \cr
&\left. + 60 k(2k -1)(\pi L)^2 c + (48 k^3 - 24 k^2 + 2k -1)(\pi L)^3 \right] ,}
}
and at $\pi L \leq c < 2 \pi L$
\eqn\eqconsii{
\eqalign{
\Re E(2 \pi L k + c) = &- {16 \over 45} G_N D_0^2 L^2 \pi^4(k + 1)(2k + 1)
\left[30 c^3 + 15 (8k -3)\pi L c^2\right. \cr
&\left.+ 60k(2k -3)(\pi L)^2 c + (48 k^3 -72 k^2 +74 k + 15)(\pi L)^3\right] .}
}
From Eqs.\eqconsi\ and \eqconsii\ the behavior of the real part of the self energy
is shown in figs.4(a) and 4(b). 
The magnified plot of fig.4(a) for the range $0<R<2\pi L$ is given in fig.4(b).
\bigskip
\vbox{
\centerline{\epsfbox{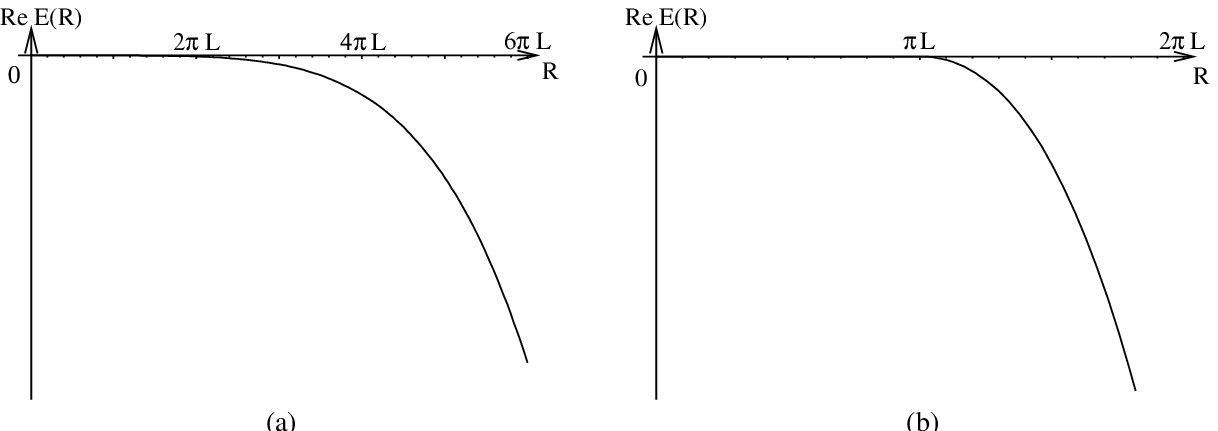}}
\vskip .5cm
\centerline{\fig\figi{The real part of the self energy.}\ 
The real part of the self energy $(\rho = D_0)$.
The magnified plot is shown in (b).}}
\bigskip

Since, substituting $k = 0$ in Eq.\eqconsi , we obtain $\Re E(R) = 0$,  
we again conclude that for $0 < R\leq \pi L$ the gravitational force of ordinary 
graviton exchange is ``screened" by the coherent effect of tachyonic graviton exchange of 
the Kaluza-Klein mode tower.

\newsec{Discussions}
We first discuss the generic feature of the gravitational screening effect due to 
the tachyonic Kaluza-Klein modes. 
Let us note that the summation formulas, Eqs.(A.1) and (A.2), give the identity:
\eqn\scr{  i\sum_{n=1}^{\infty}{1\over n}f_n(r,l)={l\over L}
-i\log{{\sin{{r+l}\over 2L}}\over{\sin{{r-l}\over 2L}}}\ , \quad 0<r-l\leq  r+l<2\pi L . }
Actually the infinite sum on the left-hand side is a periodic function 
in $r+l$ and $r-l$ with period $2\pi L$. 
So, outside the range $0<r-l\leq r+l<2\pi L$ the identity should be modified properly 
to hold by taking into account the periodicity so that it
matches the valid range in  $x$ of 
Eqs.(A.1) and (A.2). 
This point is important in calculating the real part of the gravitational self energy 
for a larger value of radius $R$ in the range $R\geq \pi L$.
Now, with the use of the formula \scr\ in Eqs.\eqt, \eqo\ and \eqene\ we find exactly
\eqn\scrng{  E(R)=-i16\pi^2G_NL \int_0^R dr \int_0^r dl\ \rho(r) \rho(l) r l 
\log{\sin{{r+l}\over 2L}\over\sin{{r-l}\over 2L}} \ ,
\quad   0<R\leq \pi L ,}
which proves that the real part of the self energy vanishes identically for the range 
$0<R\leq \pi L$ 
for any spherical mass density $\rho(r)$. 
We can therefore conclude generically that in the corresponding region the gravitational force 
due to the massless graviton exchange is ``screened" 
by the effect of the tachyonic graviton exchange 
of the Kaluza-Klein modes.

Next we discuss  a possible gravitational stability of a spherical body 
with critical radius $R$ being  equal to some  positive integer multiple of $2\pi L$. 
As presented in the  previous section,  the spherical body with the 
mass density $\rho(r)=D/r$ can be gravitationally stable at the critical radius of R equal to every 
positive integer multiple of $2\pi L$. More generally, we can 
consider an onion-like hybrid model of spherical mass density with two kinds of the $D$ value. 
We assume $\rho(r)$ to be 
\eqn\hybrid
{\rho(r)=\cases{  {D\over r}    & for  $\max[0, (2km-1)\pi L]<r<(2km+1)\pi L$         \cr
              {bD\over r}  & for  $(2km+1)\pi L<r<{\big(2(k+1)m-1\big)}\pi L$    \cr } ,
\quad  k=0, 1, 2, \cdots 
 } 
with a fixed positive integer $m$ and a positive constant $b\neq 1$ . 
Then the imaginary part of the self energy $E(R)$ is found to vanish exactly 
at critical values of $R=k(2m\pi L)$ 
with $k=1, 2, 3, \cdots$ and, 
in between any two consecutive critical values of $R$ at distance $2m\pi L$,  
it oscillates with a pitch of $2\pi L$ 
between the two values of weight proportional to $D^2$  and $ (1-b)^2 D^2$. 
When $b=1$, the onion-like hybrid model just reduces to the simple case of subsection {\it 2.1}, 
while by varying $b$ and $m$ we find to obtain a variety of the hybrid models 
with the common generic feature of the gravitational stability which shows up 
in each model at critical radius $R=2\pi Lp$ with a corresponding positive integer $p$.

\newsec{Conclusions}
We considered one extra timelike dimension, where only gravitons propagate, 
and its compactification on a circle of a radius $L$, 
which leads to the tachyonic Kaluza-Klein modes of gravitons.
And we calculated the gravitational self energies of spherical bodies of a radius
$R$ which include the contributions of the tachyonic gravitons 
as well as of a massless graviton.

We discussed two typical models of  mass density. 
At first we set the density as the spherically symmetric one,   
$\rho(r) = {D \over r}$. The imaginary part of the self energy has a 
periodicity of $2 \pi L$ in $R$ . Since this imaginary part vanishes 
at $R = 2 \pi L k$ for any $k \in {\bf N}$, the spherical body of a size 
$R=2 \pi L k$ is gravitationally stable. On the other hand the real part of 
the self energy becomes zero for $0 < R \leq \pi L$. So the gravitational force is 
screened in the spherical body with a size $0 < R \leq \pi L$. 

Next we set $\rho(r) = D_0$ (constant). The imaginary part of the self energy
is not zero for any $R$, and there is no stable size of spherical bodies.
This result is different from the one of $\rho(r) = {D \over r}$. On the other
hand, since the real part of the self energy vanishes for $0 < R \leq \pi L$, 
the gravitational force is screened in the constant spherical body 
with a size $0 < R \leq \pi L$. 
This behavior is the same as the one of $\rho(r) = {D \over r}$. 
In fact we have shown that the gravitational screening due to tachyonic gravitons 
for the range $0<R\leq\pi L$ is the generic feature of the timelike Kaluza-Klein modes 
for any spherical mass density.

Though we have considered one extra timelike dimension, there is 
no reason why no more extra timelike dimensions exist. 
Actually even in the case of any additional extra timelike dimensions we are able to calculate 
the gravitational self energies of spherical bodies in some approximation 
and to discuss their stability and screening effect in a convincing way based 
on the obtained analytic results \ref\SS{S.Matsuda and S.Seki, {\it  in preparation}.}.

As a final comment we raise the possibility that what  we have presented in this paper, 
that is,  
the fact that the scale of critical radius $R$ in the spacelike dimensions typically 
represented by a massive particle size is simply related to the scale $2\pi L$  
of the timelike dimension through the gravitational stability condition  given by 
the vanishing imaginary part of the self-energy might involve some implications 
in the construction of a final theory of particle physics 
like superstring theory or brane theory.

We also add that extra timelike dimensions have been considered in various contexts
\ref\CK{M. Chaichian and A. B. Kobakhidze, ``Mass hierarchy and localization of gravity in extra time", 
hep-th/0003269, and references cited therein.}.

\bigbreak\bigskip\bigskip\centerline{{\bf Acknowledgments}}\nobreak
This work is supported in part by the Grant-in-Aid for Scientific Research on Priority Area 707 
``Supersymmetry and Unified Theory of Elementary Particles", Japan Ministry of Education.  
S. M. is also funded partially by the Grant-in-Aid for Scientific Research (C) (2)-10640260, while  
S. S. is supported in part by JSPS Research Fellowship for Young Scientists.

\appendix{A}{}
The following infinite summations are known to hold:
\eqnn\ai
\eqnn\aii
$$
\eqalignno{
&\sum_{n=1}^\infty {\sin nx \over n} = {1 \over 2}(\pi - x),\quad 0 < x < 2\pi, &\ai \cr
&\sum_{n=1}^\infty {\cos nx \over n} = -\log\left(2\sin{x \over 2}\right),\quad 0 < x < 2\pi, &\aii
}$$

\listrefs
\bye